\documentclass[conference]{IEEEtran}
\usepackage[utf8]{inputenc}

\usepackage[T1]{fontenc}
\usepackage[normalem]{ulem}
\usepackage[english,french]{babel}

\usepackage{verbatim}
\usepackage{graphicx}
\usepackage{latexsym}

\usepackage{graphics}
\usepackage{epsfig}
\usepackage{amsmath}
\usepackage{rotating}
\usepackage{url}
\usepackage{fancyhdr}

\usepackage{tabularx}
\usepackage{array}
\usepackage{amsthm}
\usepackage{amsmath}
\usepackage{amssymb}
\usepackage{makeidx}
\usepackage{multirow}

\usepackage{multicol}
\usepackage{url}
\usepackage{textcomp}
\usepackage{cite}
\usepackage{float}

\usepackage{tikz}
\usepackage{pstricks}
\usepackage{caption}

\usepackage{blindtext}
\usepackage{etoolbox}
\usepackage{pifont} 

\usetikzlibrary{trees}

\theoremstyle{definition}
\newtheorem{thm}{Theorem}[section]

\newtheorem{pre}[thm]{Proposition}

\newtheorem{defini}[thm]{Definition}
\newtheorem{exemple}[thm]{Example}

\ifCLASSINFOpdf
\else
\fi
\usepackage{url}
\begin{document}
\selectlanguage{english}
\setlength{\columnseprule}{0pt}
\title{Secrecy by Witness-Functions on Increasing Protocols}

\author{\IEEEauthorblockN{Jaouhar Fattahi}
\IEEEauthorblockA{
Computer Science Department\\Université Laval\\Québec, QC, Canada\\
jaouhar.fattahi.1@ulaval.ca}

\and
\IEEEauthorblockN{Mohamed Mejri}
\IEEEauthorblockA{
Computer Science Department\\Université Laval\\Québec, QC, Canada\\
 mohamed.mejri@ift.ulaval.ca
}

\and
\IEEEauthorblockN{Hanane Houmani}
\IEEEauthorblockA{
Computer Science Department\\University Hassan II\\Casablanca, Morocco\\
hanane.houmani@ift.ulaval.ca 
}
%\and
%\IEEEauthorblockN{James Kirk\\ and Montgomery Scott}
%\IEEEauthorblockA{Starfleet Academy\\
%San Francisco, California 96678-2391\\
%Fax: (888) 555--1212}
}


%Telephone: (800) 555--1212, Fax: (888) 555--1212}



\maketitle

\begin{abstract}

In this paper, we present a new formal method to analyze cryptographic protocols statically for the property of secrecy. It consists in inspecting the level of security of every component in the protocol and making sure that it does not diminish during its life cycle. If yes, it concludes that the protocol keeps its secret inputs. We analyze in this paper an amended version of the Woo-Lam protocol using this new method. \\

\end{abstract}

\textit{\textbf{Keywords- Analysis; Cryptographic protocols; Secrecy.}}
%%\setlength{\footnotesep}{0.55cm}
%% \vspace*{-1cm}
%%\blfootnote{Manuscript received August 4, 2014. }
%%\blfootnote{J. Fattahi is with LSI, Computer Science and Software Engineering Department, Laval University. Adrien-Pouliot 
%%1065, Avenue of Medicine. Quebec. G1V 0A6. Canada. QC. CA (e-mail: jaouhar.fattahi.1@ulaval.ca).}
%%\blfootnote{M. Mejri is with LSI, Computer Science and Software Engineering Department, Laval University. Adrien-Pouliot 
%%1065, Avenue of Medicine. Quebec. G1V 0A6. Canada. QC. CA (e-mail: mohamed.mejri@ift.ulaval.ca).}
%%\blfootnote{M. Miraoui is with LaTIS, Department of Electrical Engineering, School of Superior Technologies (É.T.S), 1100, Notre-Dame Ouest, Montreal. H3C 1K3. Canada. QC. CA (e-mail: moeiz.miraoui.1@ens.etsmtl.ca).}
%%\blfootnote{H. Houmani is with ENSEM, University Hassan II, Casablanca, Marocco.(e-mail: hanane.houmani@ift.ulaval.ca).}
\normalsize

% \fi
\IEEEpeerreviewmaketitle

%%\marginnote{This is a margin note using the geometry package, set at 0cm vertical offset to the first line it is typeset.}[0cm]

\section{INTRODUCTION}

In this paper, we present the witness-functions as a new formal method for analyzing protocols and we run an analysis on an amended version of the Woo-Lam protocol using one of them. The Witness-Functions have been recently introduced by Fattahi et al. ~\cite{WitnessArt1,WitnessArt2,2014arXiv1408.2774F,WitnessArt3,Fatt1407:Relaxed} to statically analyze cryptographic protocols for secrecy. A protocol analysis with a witness-function consists in inspecting every component in the protocol in order to make sure that its security never drops between any receiving step and a subsequent sending one. If yes, the protocol is said to be increasing and we conclude that it keeps its secret inputs. We use the witness-function to evaluate the security of every component in the protocol. 

This paper is organized as follows:
\begin{itemize}
\item First, we give some notations that we will use in this paper;
\item then, in the section \ref{sectionPreuveThFond}, we give some abstract conditions on a function to be safe for a protocol analysis and we state that an increasing protocol keeps its secret inputs when analyzed using such functions;
\item then, in the sections \ref{sectionFonctionsetSelections} and \ref{sectionWF}, we present the witness-function and we highlight its advantages, particularly its static bounds. We state the theorem of protocol analysis with the witness-functions, as well;
\item then, in the section \ref{sectionAnWL2}, we run an analysis on an amended version of the Woo-Lam protocol and we interpret the results;
\item finally, we compare our witness-functions with some related works and we conclude. 
\end{itemize}

\section*{ Notations}
Here, we give some notations and conventions that will be used throughout the paper. 
\begin{itemize}
\item[+] We denote by ${\cal{C}}=\langle{\cal{M}},\xi,\models,{\cal{K}},{\cal{L}}^\sqsupseteq,\ulcorner.\urcorner\rangle$ the context containing the parameters that affect the analysis of a protocol:
\begin{itemize}
\item[$\bullet$] ${\cal{M}}$: is a set of messages built from the algebraic signature $\langle\cal{N}$,$\Sigma\rangle$ where ${\cal{N}}$ is a set of atomic names (nonces, keys, principals, etc.) and $\Sigma$ is a set of functions ($enc$:\!: encryption\index{Encryption}, $dec$:\!: decryption\index{Décryption}, $pair$:\!: concatenation (denoted by "." here), etc.). i.e. ${\cal{M}}=T_{\langle{\cal{N}},\Sigma\rangle}({\cal{X}})$. We use $\Gamma$ to denote the set of all substitution from $ {\cal{X}} \rightarrow {\cal{M}}$.
We designate by $\cal{A}$ all atomic messages (atoms)  in ${\cal{M}},$ by ${\cal{A}}(m)$ the set of atomic messages in $m$ and by ${\cal{I}}$ the set of principals including the intruder $I$. We denote by $k{^{-1}}$ the reverse key of a key $k$ and we consider that $({k^{-1}})^{-1}=k$.
\item[$\bullet$] $\xi$: is the theory that describes the algebraic properties of the functions in $\Sigma$ by equations. e.g. $dec(enc(x,y),y^{-1})=x$. 
\item[$\bullet$] $\models$: is the inference system of the intruder under the theory. Let $M$ be a set of messages and $m$ a message. $M$ $\models$ $m$ designates that the intruder is able to infer $m$ from $M$ using her capacity. We extrapolate this notation to traces as following: $\rho$ $\models$ $m$ designates that the intruder can infer $m$ from the messages of the trace $\rho$.
\item[$\bullet$] ${\cal{K}}$ : is a function from ${\cal{I}}$ to ${\cal{M}}$, that assigns to any principal a set of atomic messages describing her initial knowledge. We denote by $K_{{\cal{C}}}(I)$ the initial knowledge of the intruder, or simply $K(I)$ where the context is obvious.
\item[$\bullet$] ${\cal{L}}^\sqsupseteq$ : is the security lattice $({\cal{L}},\sqsupseteq, \sqcup, \sqcap, \bot,\top)$ used to assign security values to messages. 
A concrete example of a lattice is $ (2^{\cal{I}},\subseteq,\cap,\cup,\cal{I}, \emptyset)$ that will be used in this paper. 
\item[$\bullet$] $\ulcorner .\urcorner$ : is a partial function that assigns a value of security (type) to a message in ${\cal{M}}$. Let $M$ be a set of messages and $m$ a sigle message. We write $\ulcorner M \urcorner \sqsupseteq \ulcorner m \urcorner$ when
$\exists m' \in M. \ulcorner m' \urcorner \sqsupseteq \ulcorner m \urcorner$
%% \item[$\bullet$] 
\end{itemize}
\item[+] Let $p$ be a protocol, we denote by $R_G(p)$ the set of the generalized roles extracted from $p$. A generalized role is an abstraction of the protocol where the emphasis is put on a specific principal and all the unknown messages are replaced by variables. More details about the role-based specification could be found in~\cite{Debbabi11,Debbabi22,Debbabi33}.
We denote by ${\cal{M}}_p^{\cal{G}}$ the set of messages (closed and with variables) generated by $R_G(p)$, by ${\cal{M}}_p$ the set of closed messages generated by substitution in terms in ${\cal{M}}_p^{\cal{G}}$. We denote by $R^-$ (respectively $R^+$) the set of received messages (respectively sent messages) by a principal in the role $R$. Conventionally, we use uppercases for sets or sequences and lowercases for single elements. For example $M$ denotes a set of messages, $m$ a message, $R$ a role composed of sequence of steps, $r$ a step and $R.r$ the role ending by the step $r$.
\item[+] A valid trace is a close message obtained by substitution in the generalized roles. We denote by $ [\![p]\!]$ the infinite set of valid traces of $p$.
\item[+] We suppose that the intruder has the full-control of the net as given in the Dolev-Yao model~\cite{DolevY1}. We assume no restriction neither on the size of messages nor on the number of sessions.
\end{itemize}

\section{An Increasing Protocol Keeps Its Secret Inputs}\label{sectionPreuveThFond}

Hereafter, we give two abstract conditions on a function to be good for verification (safe). Then, we enunciate that an increasing protocol keeps its secret inputs.

\subsection{Safe Functions}
\begin{defini}{(Well-built Function)}\label{bienforme}
{
Let ${F}$ be a function and ${\mathcal{C}}$ be a context. ${F}$ is ${\mathcal{C}}$-well-built iff:
$\forall M,M_1,M_2 \subseteq {\mathcal{M}}, \forall \alpha \in {\mathcal{A}}({\mathcal{M}}) \mbox{:}$
$
\left\{
\begin{array}{ll}
{F}(\alpha,\{\alpha\})= \bot; & \\
{F}(\alpha, {M}_1 \cup {M}_2)= {F}(\alpha, {M}_1)\sqcap{F}(\alpha,{M}_2); & \\
{F}(\alpha,{M}) =\top, \mbox{ if } \alpha \notin {\mathcal{A}}({M}). & \\
\end{array}
\right.
$
}
\end{defini}

A well-built function ${F}$ must return the infimum for an atom $\alpha$ that appears in clear in $M$ to express the fact that is exposed to everybody in $M$. It should return for it in the union of two sets, the minimum of the two values evaluated in each set apart. It returns the supremum for any atom $\alpha$ that does appear in $M$ to express the fact that none could deduce it from $M$.

\begin{defini}{(Invariant-by-Intruder Function)}\label{spi}
{
Let ${F}$ be a function and ${\mathcal{C}}$ be a context. 
${F}$ is ${\mathcal{C}}$-invariant-by-intruder iff:\\
$
\forall {M} \subseteq {\mathcal{M}}, m\in {\mathcal{M}}. {M} \models_{\mathcal{C}} m \Rightarrow \forall \alpha \in {\mathcal{A}}(m). ({F}(\alpha,m) \sqsupseteq{F}(\alpha,{M})) \vee (\ulcorner K(I) \urcorner \sqsupseteq \ulcorner \alpha \urcorner).
$
}
\end{defini}

An invariant-by-intruder function ${F}$ is such that, when it assigns a security value to an atom $\alpha$ in a set of messages $M$ the intruder can never deduce, using her knowledge, from $M$ another message $m$ in which this value decreases (i.e. ${F}(\alpha,m) \not \sqsupseteq{F}(\alpha,{M})$), except when $\alpha$ is intentionally destined to the intruder (i.e. $\ulcorner K(I) \urcorner \sqsupseteq \ulcorner \alpha \urcorner$).

\begin{defini}{(Safe Function)}
{
Let ${F}$ be a function and ${\mathcal{C}}$ be a context.
\[{F} \mbox { is }{\mathcal{C}}\mbox{-safe } \mbox{ iff } \left\{
\begin{array}{ll}
{F} \mbox{ is } {\mathcal{C}}\mbox{-well-built} &
\\
{F} \mbox{ is } {\mathcal{C}}\mbox{-invariant-by-intruder}& 
\end{array}
\right.
\]
}
\end{defini} 
A safe function ${F}$ is well-built and invariant-by-intruder.

\begin{defini}{(${F}$-Increasing Protocol)}\label{ProAbsCroi}
{
Let ${F}$ be a function, ${\mathcal{C}}$ be a context and $p$ be a protocol.\\
$p$ is ${F}$-increasing in ${\mathcal{C}}$ iff:\\
$\forall R.r \in R_G(p),\forall \sigma \in \Gamma: {\mathcal{X}} \rightarrow {\mathcal{M}}_p \mbox{ we have: }$
\[
\forall \alpha \in {\mathcal{A}}({\mathcal{M}}).{F}(\alpha, r^+\sigma)\sqsupseteq \ulcorner \alpha \urcorner \sqcap{F}(\alpha, R^-\sigma)
\]
}
\end{defini}

An ${F}$-increasing protocol generates permanently traces with atomic messages having always a security value, evaluated by ${F}$, higher when sending (i.e. in $ r^+\sigma$) than it was on its reception (i.e. in $R^-\sigma$).

\begin{thm}{(Security of Increasing Protocols)}\label{mainTh}
{
Let ${F}$ be a ${\mathcal{C}}$-safe Function and $p$ an ${F}$-increasing protocol.
\begin{center}
$p$ keeps its secret inputs.
\end{center}
}
\end{thm}

The theorem \ref{mainTh} states that a protocol is secure when verified by a safe function $F$ on which it is proved increasing. That is, if the intruder manages to infer a secret $\alpha$ (get it in clear), then its value returned by $F$ is the infimum because $F$ is well-built. That could not happen due to the protocol rules because the protocol is increasing by $F$ unless $\alpha$ has initially the infimum. In this case, $\alpha$ was not from the beginning a secret. That could not happen neither by using the capacity of the intruder because $F$ is invariant-by-intruder. Therefore, the secret is kept forever.

\section{Safe Functions}\label{sectionFonctionsetSelections}\index{Fonction d'interprétation}

Now, we define three practical functions that meet the conditions or safety: $F_{MAX}^{EK}$, $F_{N}^{EK}$ and $F_{EK}^{EK}$. Each function among them returns for an atom $\alpha$ in a message $m$:
\begin{enumerate}
\item if $\alpha$ is encrypted by a key $k$, where $k$ is the most external protective key (shortly the external protective key denoted by EK) that satisfies: $\ulcorner k^{-1} \urcorner \sqsupseteq \ulcorner \alpha \urcorner$, \textit{any subset} among the principals that know $k^{-1}$ and the principals that travel with $\alpha$ under the same protection by $k$.
At this step:
\begin{enumerate}
\item $F_{MAX}^{EK}$ returns the set of all these candidates;
\item $F_{N}^{EK}$ returns the set of principals that travel with $\alpha$ under the same protection by $k$;
\item $F_{EK}^{EK}$ returns the set of principals that know $k^{-1}$.
\end{enumerate}
\item for two messages linked by an operator other than an encryption by a protective key (e.g. pair), the union of two values evaluated in the two messages apart by $F$.
\item if $\alpha$ does not have a protective key in $m$, the infimum to express the fact that it could be discovered by an intruder from $m$;
\item if $\alpha$ does not appear in $m$, the supremum to reflect that it could not be discovered by anybody from $m$;
\end{enumerate}

A such function is well-built by construction. It is invariant-by-intruder too. The main idea of its invariance by intruder property is that the returned candidates (principals) are selected from a section (a component of $m$) protected by $k$ (invariant by intruder). Hence, to alter this section (to lower the value of security of an atom $\alpha$), the intruder must previously have got the atomic key $k^{-1}$, so her knowledge should satisfy: $\ulcorner K(I) \urcorner \sqsupseteq \ulcorner k^{-1} \urcorner$. Since the key $ k^{-1}$ must satisfy: $\ulcorner k^{-1} \urcorner \sqsupseteq \ulcorner \alpha \urcorner$, then the knowledge of the intruder satisfy: $\ulcorner K(I) \urcorner \sqsupseteq \ulcorner \alpha \urcorner$ too (transitivity of "$\sqsupseteq$" in the lattice), which is the definition of an invariant-by-intruder function. It is very important to mention that we consider the form $m_{\downarrow}$ of a message $m$ that removes keys that cancel out (i.e. $dec(enc(m,k),k^{-1})_{\downarrow}=m$). We suppose in this paper that we do not have any other special algebraic properties in the equational theory. This will be the scope of a future work.

\begin{exemple}
{
Let $\alpha$ be an atom, $m$ be a message and $k_{ab}$ be a key such that:
$\ulcorner \alpha \urcorner=\{A, B, S\}$; $m=\{A.\{S.\alpha.D\}_{k_{as}}\}_{k_{ab}}$; $ \ulcorner{k_{ab}^{-1}}\urcorner=\{A, B\}$;\\
${F}_{MAX}^{EK}(\alpha,m)=\ulcorner{k_{ab}^{-1}}\urcorner{ \cup}\{A, S, D\}=\{A,B\} \cup \{A,S,D\}=\{A, B, S, D\}$.\\
${F}_{MAX}^{N}(\alpha,m)=\{A, S, D\}$.\\
${F}_{MAX}^{EK}(\alpha,m)=\ulcorner{k_{ab}^{-1}}\urcorner=\{A,B\}$.\\
}
\end{exemple}

In the rest of this paper $F$ refers to any of the functions $F_{MAX}^{EK}$, $F_{N}^{EK}$ and $F_{EK}^{EK}$.

\section{The witness-functions}\label{sectionWF}

According to the theorem \ref{mainTh}, if a protocol $p$ is proved $F$-increasing on its \textit{valid traces} using a safe function $F$, then it is secure. However, the set of valid traces is infinite. In order to be able to analyze a protocol from within its finite set of the generalized roles, we should adapt a safe function to the problem of substitution (variables) and look for an additional mechanism that allows us to propagate any decision made on the generalized roles to valid traces. The witness-functions are this mechanism. But first, let us introduce the derivative messages. A derivative message is a message of the generalized roles from which we exclude variables that do not contribute to the evaluation of security. This is described in the definition \ref{derivation}. 

\begin{defini}{(Derivation)}\label{derivation}
{
We define the derivative message as follows:
%%\begin{table}[h]
%% \centering
\begin{center}
\begin{tabular}{rrcll}
~~~~~~& $\partial_X \alpha$ & $=$ & $\alpha$ &\\
& $\partial_X \epsilon$ & $=$ & $\epsilon$ &\\
& $\partial_X X$ & $=$& $\epsilon$ &\\
& $\partial_X Y$ & $=$ & $Y$ &\\
& $\partial_{\{X\}} m$ & $=$ & $\partial_{X} m$ &\\
& $\partial{[\overline{X}]} m$ & $=$ & $\partial_{\{{\mathcal{X}}_m\backslash X\}} m$ &\\
& $\partial_X f(m)$ & $=$ & $ {f}(\partial_X m), f\in \Sigma$ &\\
& $\partial_{S_1 \cup S_2}m$ & $=$& $\partial_{S_1}\partial_{S_2}m$&\\
\\
%%& $\partial_{S_1 \cup S_2}m$ & $=$ & $\partial_{S_2 \cup S_1}m$&
\end{tabular}
\end{center}
}
\end{defini}

Then, we apply a safe function ${F}$ to derivative messages. For an atom in the static neighborhood (i.e. in $\partial m$), we evaluate its security with no respect to variables. Else, for any message substituting a variable, it is evaluated as a constant block, whatever its content, and with no respect to other variables, if any. This is described by the definition \ref{Fder}. 
\begin{defini}\label{Fder}
{
Let $m\in {\mathcal{M}}_p^{\mathcal{G}}$, $X \in {\mathcal{X}}_m$ and $m\sigma$ be a valid trace. 
For all $\alpha \in {\mathcal{A}}(m\sigma)$, $\sigma\in\Gamma$, we denote by:
\[
{F}(\alpha, \partial [\overline{\alpha}] m\sigma) = \left\{
\begin{array}{ll}
{F}(\alpha,\partial m) & \mbox{if } \alpha \in {\mathcal{A}}(\partial m),\\
{F}(X,\partial [\overline{X}] m) & \mbox{if }\alpha \notin {\mathcal{A}}(\partial m) \\
& \mbox{and } \alpha =X\sigma.
\end{array}
\right.
\]
}
\end{defini}

The application in the definition \ref{Fder} could not be used to analyze protocols. It is harmful. Let us examine its deficiency in the example \ref{exempleApp}.
\begin{exemple}\label{exempleApp}
{
Let $m_1$ and $m_2$ be two messages of ${\mathcal{M}}_p^{\mathcal{G}}$ such that 
$m_1=\{\alpha.D.X\}_{k_{ab}}$ and $m_2=\{\alpha.Y\}_{k_{ab}}$ and $\ulcorner\alpha\urcorner=\{A,B\}$. Let $m=\{\alpha.D.B\}_{k_{ab}}$ be in a valid trace.\\
$$F_{MAX}^{EK}(\alpha,\partial [\overline{\alpha}] m)=\begin{cases}
\{A, B, D\},& \!\!\!\!\!\mbox{if } m=m_1\sigma_1|X\sigma_1=B,\\
%% & \\
\{A, B\},& \!\!\!\!\!\!\!\!\!\!\!\mbox{if } m=m_2\sigma_2|Y\sigma_2=D.B
\end{cases}$$
Therefore, $F_{MAX}^{EK}(\alpha,\partial [\overline{\alpha}] m)$ is not a function on $m\sigma$ (i.e. it returns two possible values for the same preimage). 
}
\end{exemple}

The witness-function in the definition \ref{WF} fixes this deficiency: it looks for all the origins $m$ of the substituted message $m\sigma$ in the generalized roles, applies the application in the definition \ref{Fder} and returns the minimum that obviously exists and is unique in a lattice. 

\begin{defini}{(Witness-Function)}\label{WF}
{
Let $m\in {\mathcal{M}}_p^{\mathcal{G}}$, $X \in {\mathcal{X}}_m$ and $m\sigma$ be a valid trace.
Let $p$ be a protocol and $F$ be a ${\mathcal{C}}$-safe Function.
We define a witness-function ${\cal{W}}_{p,{F}}$ for all $\alpha \in {\mathcal{A}}(m\sigma)$, $\sigma\in\Gamma$, as follows: 
%%On définit la fonction témoin sur les messages clos comme suit: 
\[{{{\cal{W}}}}_{p,{F}}(\alpha,m\sigma)=\!\!\!\!\!\!\!\!\!\!\underset{\overset {m' \in {\mathcal{M}}_p^{\mathcal{G}}}{\exists \sigma' \in \Gamma.m'\sigma' = m \sigma }}{\sqcap}\!\!\!\!\!\!\!\!\!\! {F}(\alpha, \partial [\overline{\alpha}] m'\sigma')\]
}
\end{defini}

A witness-function ${{{\cal{W}}}}_{p,{F}}$ is safe when $F$ is. Indeed, it is easy to verify that it is well-built. It is invariant-by-intruder as well since the returned values (principal identities) are those returned by $F$ applied to derivative messages of the origins of $m\sigma$. Derivation does not add new candidates, it just removes some of them, but returns always candidates from the same invariant section by the intruder when the message is substituted.

Since the target of the witness-functions is to analyze protocols statically and since it still depends on $\sigma$ (runs), we will bind it in two static bounds and use them for analysis instead of the witness-function itself. The lemma \ref{PropBoun} provides these bounds.

\begin{pre}{(Witness-Function Bounds)}\label{PropBoun}
{
Let $m\in{\mathcal{M}}_p^{\mathcal{G}}$.
Let $F$ be a ${\mathcal{C}}$-safe function and ${\cal{W}}_{p,{F}}$ be a witness-function.
For all $\sigma \in \Gamma$ we have:
$${F}(\alpha,\partial[\overline{\alpha}] m)\sqsupseteq {{{\cal{W}}}}_{p,{F}}(\alpha,m\sigma)\sqsupseteq \!\!\!\!\!\!\!\!\!\!\!\! \underset{\overset{m' \in {\mathcal{M}}_p^{\mathcal{G}}}{\exists \sigma' \in \Gamma.m'\sigma' = m \sigma' }}{\cup} \!\!\!\!\!\!\!\!\!\!\!\!\!{F}(\alpha, \partial [\overline{\alpha}]m'\sigma')$$
}
\end{pre}

For a secret $\alpha$ in a substituted message $m\sigma$, the upper-bound ${F}(\alpha,\partial[\overline{\alpha}] m)$ evaluates its security from one confirmed origin $m$ in the generalized roles, the witness-function ${{{\cal{W}}}}_{p,{F}}(\alpha,m\sigma)$ from the set of the exact origins of $m\sigma$ (when running). The message $m$ is obviously one of them. The lower-bound $\!\!\!\!\!\!\!\!\!\!\!\! \underset{\overset{m' \in {\mathcal{M}}_p^{\mathcal{G}}}{\exists \sigma' \in \Gamma.m'\sigma' = m \sigma' }}{\cup} \!\!\!\!\!\!\!\!\!\!\!\!\!{F}(\alpha, \partial [\overline{\alpha}]m'\sigma')$ evaluates it from the set of all the messages that are unifiable with $m$. This set naturally includes the set of definition of the witness-function since unifications include substitutions. Unifications in the lower-bound trap any intrusion (odd principal identities). Please notice that both the upper-bound and the lower-bound are static (independent of $\sigma$).

\begin{thm}{(Analysis Theorem)}\label{PAT}
{
Let $p$ be a protocol. 
Le $F$ be a safe function.
Let ${\cal{W}}_{p,{F}}$ be a witness-function.
$p$ keeps its secrect inputs if:\\
$\forall R.r \in R_G(p), \forall \alpha \in {\mathcal{A}}{(r^+ )}$ we have:
$$\underset{\overset {m' \in {\mathcal{M}}_p^{\mathcal{G}}}{\exists \sigma' \in \Gamma.m'\sigma' = r^+ \sigma' }}{\sqcap} \!\!\!\!\!\!\!\!\!\! {F}(\alpha, \partial[\overline{\alpha}] m'\sigma') \sqsupseteq \ulcorner \alpha \urcorner \sqcap {F}(\alpha,\partial[\overline{\alpha}] R^-)$$
}
\end{thm}

This theorem states a static criterion for secrecy. It derives directly from the theorem \ref{mainTh} and the lemma \ref{PropBoun}. This allows us to analyze a protocol from within its generalized roles (finite set) and send any decision made-on to valid traces.

\section{Analysis of the Woo-Lam Protocol (Amended Version) with a witness-function}\label{sectionAnWL2}

Here, we analyze an amended version of the Woo-Lam protocol with a witness-function and we prove that is correct for secrecy. This version  is  denoted by $p$ in Table~\ref{WLMV:protv2p}.

          \begin{table}[h]
         \caption{Woo-Lam Protocol-Amended version}
         \label{WLMV:protv2p} 
               \begin{center}$\begin{array}{|l|}
               \hline\\
                   \begin{array}{llll}
                    p   =&\langle 1,A\rightarrow B: A\rangle.  \\
                    & \langle 2,B\rightarrow A: N_b\rangle. \\
                    & \langle 3,A\rightarrow B: \{ B.k_{ab}\}_{k_{as}}\rangle. \\
                    & \langle 4,B\rightarrow S:\{A.N_b.\{B.k_{ab}\}_{k_{as}}\}_{k_{bs}}\rangle.  \\
                    & \langle 5,S\rightarrow B:\{N_b.\{A.k_{ab}\}_{k_{bs}} \}_{k_{bs}}\rangle\\ &\\
                    \end{array} \\ \hline  \end{array}$
         \end{center}

         \end{table}

    The   role-based   specification  of $p$  is ${\cal R}_G(p) = \{{\cal A}_G ^1,
    ~{\cal  A}_G  ^2,~{\cal B}_G ^1,~{\cal B}_G ^2,~{\cal B}_G ^3,~ {\cal S}_G ^1\}$,
    where the generalized roles ${\cal A}_G ^1$, ${\cal A}_G ^2$ of $A$ are as follows:
    \[\begin{array}{l}
            \begin{array}{lllll}
                    {\cal A}_G ^1 =& \langle  i.1, A    & \rightarrow & I(B):&  A
                    \rangle\\
            \end{array}\\
                    \\
             \begin{array}{llllll}
                    {\cal A}_G ^2=& \langle  i.1,  A    & \rightarrow & I(B):&  A \rangle .\\
                    & \langle i.2,  I(B) & \rightarrow & A:&  X \rangle .\\
                    & \langle i.3,  A    & \rightarrow & I(B):&  \{B.k_{ab}^i\}_{k_{as}}\rangle
             \end{array}\end{array}\]

     The generalized roles ${\cal B}_G ^1$, ${\cal B}_G ^2$, ${\cal B}_G
    ^3$  of $B$ are as follows:
            \[\begin{array}{l}
            \begin{array}{lllll}
                {\cal B}_G ^1=& \langle i.1,  I(A) & \rightarrow & B:&  A \rangle .\\
                        & \langle i.2,  B    & \rightarrow & I(A) :&  N_b^i \rangle  \\
            \end{array}\\ \\
            \begin{array}{lllll}
                    {\cal B}_G ^2=& \langle i.1,  I(A) & \rightarrow & B :&  A \rangle .\\
                    & \langle i.2,  B    & \rightarrow & I(A):&  N_b ^i\rangle .\\
                    & \langle i.3,  I(A) & \rightarrow & B:&  Y \rangle .\\
                    & \langle i.4,  B    & \rightarrow & I(S):& \{A.N_b^i.Y \}_{k_{bs}}\rangle
            \end{array}\\ \\
            \begin{array}{lllll}
                    {\cal B}_G ^3=& \langle i.1,  I(A) & \rightarrow & B:&  A \rangle .\\
                    & \langle i.2,  B    & \rightarrow & I(A):&  N_b ^i\rangle.\\
                    & \langle i.3,  I(A) & \rightarrow & B:&  Y \rangle.\\
                    & \langle i.4, B    & \rightarrow & I(S):& \{A.N_b^i.Y \}_{k_{bs}}\rangle.  \\
                    & \langle i.5,  I(S) & \rightarrow & B:&  \{N_b^i.\{A.Z\}_{k_{bs}}\}_{k_{bs}}\rangle \\
            \end{array}
             \end{array}\]

             The  generalized role $ {\cal S}_G ^1$ of $S$ is as follows:
            \[\begin{array}{l}
             \begin{array}{lllll}
                            {\cal S}_G ^1= & \langle i.4, I(B) & \rightarrow & S:& \{A.U.\{B.V\}_{k_{as}}
                            \}_{k_{bs}} \rangle. \\
                                    & \langle i.5,S & \rightarrow & I(B):& \{U.\{A.V\}_{k_{bs}}\}_{k_{bs}}\rangle\\
               \end{array}
            \end{array} \]
Let us have a context of verification such that: \\
$\ulcorner k_{as}\urcorner=\{A, S\}$; $\ulcorner k_{bs}\urcorner=\{B, S\}$; $\ulcorner k_{ab}^i\urcorner=\{A, B, S\}$; 
$\ulcorner N_b^i \urcorner=\bot$; $\forall A\in {\cal{I}}, \ulcorner A \urcorner=\bot$.\\
The principal identities are not analyzed since they are set public in the context.\\
Let $F= F_{MAX}^{EK}$; ${\cal{W}}_{p,F}= {\cal{W}}_{p,F_{MAX}^{EK}}$;\\
We denote  by ${\cal{W}}_{p,F}'(\alpha,m)$ the lower-bound $\underset{\overset {m' \in {\cal{M}}_p^{\cal{G}}}{\exists \sigma' \in \Gamma.m'\sigma' = m\sigma' }}{\sqcap}\!\!\!\!\!\!\!\!\!\!  F(\alpha, \partial [\overline{\alpha}]m'\sigma')$ of the witness-function ${\cal{W}}_{p,F}(\alpha,m)$.\\
Let  ${\cal{M}}_p^{\cal{G}}=\{
A_1,
X_1,
\{B_1.K_{A_{2}B_{1}}^i\}_{K_{A_{2}S_{1}}},
A_3 ,
N_{B_{2}}^i,
Y_1,$\\$
\{A_4.N_{B_{3}}^i.Y_2 \}_{K_{{B_{3}S_2}}},
\{N_{B_{4}}^i.\{A_5.Z_1\}_{K_{B_{4}S_3}}\}_{K_{B_{4}S_3}},\\
\{A_6.U_1.\{B_5.V_1\}_{K_{{A_6}S_4}}\}_{K_{{B_5}S_4}},
\{U_2.\{A_7.V_2\}_{K_{B_{6}S_5}}\}_{K_{B_{6}S_5}}
 \}$\\
After elimination of duplicates,
${\cal{M}}_p^{\cal{G}}=\\ \{
A_1,
X_1,
\{B_1.K_{A_{2}B_{1}}^i\}_{K_{A_{2}S_{1}}},
N_{B_{2}}^i,
\{A_4.N_{B_{3}}^i.Y_2 \}_{K_{{B_{3}S_2}}},$\\$
\{N_{B_{4}}^i.\{A_5.Z_1\}_{K_{B_{4}S_3}}\}_{K_{B_{4}S_3}},\\
\{A_6.U_1.\{B_5.V_1\}_{K_{{A_6}S_4}}\}_{K_{{B_5}S_4}},
\{U_2.\{A_7.V_2\}_{K_{B_{6}S_5}}\}_{K_{B_{6}S_5}}
 \}$\\
The variables are denoted by $X_1, Y_2, Z_1, U_1, U_2, V_1$ and $V_2$;\\ 
The static names are denoted by $A_1$, $B_1$, $K_{A_{2}B_{1}}^i$, ${K_{A_{2}S_{1}}}$, 
$N_{B_{2}}^i$, $A_4$, $N_{B_{3}}^i$, ${K_{{B_{3}S_2}}}$, 
$N_{B_{4}}^i$, $A_5$, ${K_{B_{4}S_3}}$, 
$A_6$, $B_5$, ${K_{{A_6}S_4}}$, ${K_{{B_5}S_4}}$, 
$A_7$ and ${K_{B_{6}S_5}}$.

\subsection{Analysis of the Generalized Roles of $A$}
As defined in the generalized role $A$, an agent $A$ can participate in some session $S^{i}$  in which she receives an unkown message $X$ and sends the message $\{B.k_{ab}^i\}_{k_{as}}$. This is described by the following rule: \[{S^{i}}:\frac{X}{\{B.k_{ab}^i\}_{k_{as}}}\]

\noindent{-Analysis of the messages exchanged  in $S^{i}$:}\\
\\
1- For any $k_{ab}^i$:\\
\\
a- When receiving: $R_{S^{i}}^-=X$ \textit{(on receiving, we use the upper-bound)}\\
$F(k_{ab}^i,\partial  [\overline{k_{ab}^i}]X )=F(k_{ab}^i,\epsilon)=\top$ (1.0)\\
\\
b- When sending: $r_{S^{i}}^+=\{B.k_{ab}^i\}_{k_{as}}$ \textit{(on sending, we use the lower-bound)}\\
 $\forall k_{ab}^i.\{m' \in {\cal{M}}_p^{\cal{G}}|{\exists \sigma' \in \Gamma.m'\sigma' = r_{S^{i}}^+\sigma' } \}$\\$=\forall k_{ab}^i.\{m' \in {\cal{M}}_p^{\cal{G}}|{\exists \sigma' \in \Gamma.m'\sigma' = \{B.k_{ab}^i\}_{k_{as}}\sigma' } \}$ \\$=\{(\{B_1.K_{A_{2}B_{1}}^i\}_{K_{A_{2}S_{1}}},\sigma_1')\}$ such that: $ \sigma_1'=\{ B_1 \longmapsto B, K_{A_{2}B_{1}}^i \longmapsto k_{ab}^i, {K_{A_{2}S_{1}}} \longmapsto {k_{as}}\}$\\
${\cal{W}}_{p,F}'(k_{ab}^i,\{B.k_{ab}^i\}_{k_{as}})$\\
$=\{\mbox{Definition of the lower-bound of the witness-function}\}$\\
$F(k_{ab}^i,\partial[\overline{k_{ab}^i}] \{B_1.K_{A_{2}B_{1}}^i\}_{K_{A_{2}S_{1}}} \sigma_{1}')$\\
$=\{\mbox{Extracting the static neighborhood}\}$\\
$F(k_{ab}^i,\partial[\overline{k_{ab}^i}] \{B.k_{ab}^i\}_{k_{as}} \sigma_{1}')$ \\
$=\{\mbox{Definition } \ref{Fder}\}$
\\
$F(k_{ab}^i,\partial[\overline{k_{ab}^i}] \{B.k_{ab}^i\}_{k_{as}})$ \\
%%$=\{\mbox{Since } F=F_{MAX}^{EK}\}$\\
$=\{\mbox{Derivation in the definition } \ref{derivation}\}$
\\
$F(k_{ab}^i,\{B.k_{ab}^i\}_{k_{as}})$ \\
$=\{\mbox{Since } F=F_{MAX}^{EK}\}$\\
$\{B, A, S\}$(1.1)\\
\\
2- Compliance with the theorem \ref{PAT}:\\
From (1.0) and (1.1), we have: ${\cal{W}}_{p,F}'(k_{ab}^i,\{B.k_{ab}^i\}_{k_{as}})= \{A, B, S\} \sqsupseteq \ulcorner k_{ab}^i \urcorner \sqcap F(k_{ab}^i,\partial  [\overline{k_{ab}^i}]X )=\{A, B, S\}$ (1.2)\\
From (1.2) we have: the messages exchanged in the session $S^{i}$ (i.e. $k_{ab}^i$) respect the theorem \ref{PAT}. (I)

\subsection{Analysis of the generalized roles of $B$}
As defined in the generalized roles of $B$, an agent $B$ can participate in two subsequent sessions: $S^{i}$ and $S^{j}$ such that $j>i$. In the former session $S^{i}$, the agent $B$ receives the identity $A$ and sends the nonce $N_b ^i$. In the subsequent session $S^{j}$, she receives an unknown message $Y$ and she sends the message $\{A.N_b^i.Y \}_{k_{bs}}$. This is described by the following rules:
\[{S^{i}}:\frac{A}{N_b ^i} ~~~~~~~~~~~~~~~~~~~~~~~~ {S^{j}}:\frac{Y}{\{A.N_b^i.Y \}_{k_{bs}}}\]
$ $\\
\noindent{-Analysis of the messages exchanged  in  $S^{i}$:}\\
\\
1- For any $N_b^i$:\\
Since $N_b^i$ is declared public in the context (i.e. $\ulcorner N_b^i \urcorner=\bot$), then we have directly:\\
${\cal{W}}_{p,F}'(N_b^i,N_b^i)\sqsupseteq \ulcorner N_b^i \urcorner \sqcap F(N_b^i,\partial  [\overline{N_b^i}]A )=\bot$ (2.1)\\
\\
\noindent{-Analysis of the messages exchanged  in $S^{j}$:}\\
\\
1- For any $N_b^i$:\\
Since $N_b^i$ is declared public in the context (i.e. $\ulcorner N_b^i \urcorner=\bot$), then we have directly:\\
${\cal{W}}_{p,F}'(N_b^i,\{A.N_b^i.Y \}_{k_{bs}})\sqsupseteq \ulcorner N_b^i \urcorner \sqcap  F(N_b^i,\partial  [\overline{N_b^i}]Y )=\bot$ (2.2)\\
\\
2- For any $Y$:\\
Since when receiving, we have $F(Y,\partial  [\overline{Y}]Y )=F(Y,Y)=\bot$, then we have directly:\\
${\cal{W}}_{p,F}'(Y,\{A.N_b^i.Y \}_{k_{bs}})\sqsupseteq \ulcorner Y \urcorner \sqcap F(Y,\partial  [\overline{Y}]Y )=\bot$ (2.3)\\
\\
3- Compliance with the theorem \ref{PAT}:\\
From (2.1), (2.2) and (2.3) we have: the messages exchanged in the session $S^{i}$ and $S^{j}$  respect the theorem \ref{PAT}. (II)

\subsection{Analysis of the generalized roles of $S$}
As defined in the generalized role $S$, an agent $S$ can participate in some session $S^{i}$  in which she receives the message $ \{A.U.\{B.V\}_{k_{as}}\}_{k_{bs}}$ and sends the message $ \{U.\{A.V\}_{k_{bs}}\}_{k_{bs}}$. This is described by the following rule: \[{S^{i}}:\frac{\{A.U.\{B.V\}_{k_{as}}\}_{k_{bs}}}{ \{U.\{A.V\}_{k_{bs}}\}_{k_{bs}}}\]
1- For any $U$:\\
\\
b- When receiving: $R_{S^{i}}^-=\{A.U.\{B.V\}_{k_{as}}\}_{k_{bs}}$ \textit{(on receiving, we use the upper-bound)}\\
$F(U,\partial  [\overline{U}]\{A.U.\{B.V\}_{k_{as}}\}_{k_{bs}} )=$\\$F(U,\{A.U.\{B\}_{k_{as}}\}_{k_{bs}} )=\{A, B, S\}$ (3.2)\\
\\
b-When sending: $r_{S^{i}}^+=\{U.\{A.V\}_{k_{bs}}\}_{k_{bs}}$ \textit{(on sending, we use the lower-bound)}\\
 $\forall U.\{m' \in {\cal{M}}_p^{\cal{G}}|{\exists \sigma' \in \Gamma.m'\sigma' = r_{S^{i}}^+\sigma' } \}$\\$=\forall U.\{m' \in {\cal{M}}_p^{\cal{G}}|{\exists \sigma' \in \Gamma.m'\sigma' = \{U.\{A.V\}_{k_{bs}}\}_{k_{bs}} \sigma' } \}$ \\$=\{(\{\{U_2.\{A_7.V_2\}_{K_{B_{6}S_5}}\}_{K_{B_{6}S_5}},\sigma_1')\}$ such that: $\sigma_1'=\{ U_2 \longmapsto U, A_7\longmapsto A, V_2 \longmapsto V, {K_{B_{6}S_5}}\longmapsto  {k_{bs}}\}$\\
\\
${\cal{W}}_{p,F}'(U,\{U.\{A.V\}_{k_{bs}}\}_{k_{bs}})$\\
$=\{\mbox{Definition of the lower-bound of the witness-function}\}$\\
$F(U,\partial[\overline{U}]\{U_2.\{A_7.V_2\}_{K_{B_{6}S_5}}\}_{K_{B_{6}S_5}} \sigma_{1}') $\\
$=\{\mbox{Extracting the static neighborhood}\}$\\
$F(U,\partial[\overline{U}] \{U.\{A.V\}_{k_{bs}}\}_{k_{bs}} \sigma_{1}')$ \\
$=\{\mbox{Definition } \ref{Fder}\}$
\\
$F(U,\partial[\overline{U}] \{U.\{A.V\}_{k_{bs}}\}_{k_{bs}})$ \\
%%$=\{\mbox{Since } F=F_{MAX}^{EK}\}$\\
$=\{\mbox{Derivation in the definition } \ref{derivation}\}$
\\
$F(U,\{U.\{A\}_{k_{bs}}\}_{k_{bs}})$ \\
$=\{\mbox{Since } F=F_{MAX}^{EK}\}$\\
$\{A, B, S\}$(3.2)\\
\\
2- For any $V$:\\
\\
a- When receiving: $R_{S^{i}}^-=\{A.U.\{B.V\}_{k_{as}}\}_{k_{bs}}$ \textit{(on receiving, we use the upper-bound)}\\
$F(V,\partial  [\overline{V}]\{A.U.\{B.V\}_{k_{as}}\}_{k_{bs}})=$\\$F(V,\{A.\{B.V\}_{k_{as}}\}_{k_{bs}} )=$ \\$\left\{
    \begin{array}{ll}
        \{A, B, S\} & \mbox{if } k_{as} \mbox{ is the external protective key} \\
                          & \mbox{of } V \mbox{ in } \{A.\{B.V\}_{k_{as}}\}_{k_{bs}} \\
                          \\
        \{A, B, S\} & \mbox{if } k_{bs} \mbox{ is the external protective key} \\
                          & \mbox{of } V \mbox{ in } \{A.\{B.V\}_{k_{as}}\}_{k_{bs}} 
   \end{array}=
\right.$ $\{A, B, S\}$
 (3.3)\\
\\
b-When sending: $r_{S^{i}}^+=\{U.\{A.V\}_{k_{bs}}\}_{k_{bs}}$ \textit{(on sending, we use the lower-bound)}\\
 $\forall V.\{m' \in {\cal{M}}_p^{\cal{G}}|{\exists \sigma' \in \Gamma.m'\sigma' = r_{S^{i}}^+\sigma' } \}$\\$=\forall V.\{m' \in {\cal{M}}_p^{\cal{G}}|{\exists \sigma' \in \Gamma.m'\sigma' = \{U.\{A.V\}_{k_{bs}}\}_{k_{bs}} \sigma' } \}$ \\$=\{(\{\{U_2.\{A_7.V_2\}_{K_{B_{6}S_5}}\}_{K_{B_{6}S_5}},\sigma_1'),$\\$(\{N_{B_{4}}^i.\{A_5.Z_1\}_{K_{B_{4}S_3}}\}_{K_{B_{4}S_3}},\sigma_2')\}$ such that:
$$\left\{
    \begin{array}{l}
      \sigma_1'=\{ U_2 \longmapsto U, A_7\longmapsto A, V_2 \longmapsto V, {K_{B_{6}S_5}}\longmapsto  {k_{bs}}\}   \\
      \sigma_2'=\{ U \longmapsto N_{B_{4}}^i, A_5\longmapsto A, Z_1 \longmapsto V, {K_{B_{4}S_3}}\longmapsto  {k_{bs}}\}  \\
    \end{array}
\right.$$
${\cal{W}}_{p,F}'(V,\{U.\{A.V\}_{k_{bs}}\}_{k_{bs}})$\\
$=\{\mbox{Definition of the lower-bound of the witness-function}\}$\\
$F(V,\partial[\overline{V}]\{U_2.\{A_7.V_2\}_{K_{B_{6}S_5}}\}_{K_{B_{6}S_5}} \sigma_{1}') \sqcap$\\$ F(V,\partial[\overline{V}]\{N_{B_{4}}^i.\{A_5.Z_1\}_{K_{B_{4}S_3}}\}_{K_{B_{4}S_3}} \sigma_{2}') $\\
$=\{\mbox{Extracting the static neighborhood}\}$\\
$F(V,\partial[\overline{V}] \{U.\{A.V\}_{k_{bs}}\}_{k_{bs}} \sigma_{1}')\sqcap$\\$ F(V,\partial[\overline{V}] \{ N_{B_{4}}^i.\{A.V\}_{k_{bs}}\}_{k_{bs}} \sigma_{2}')$ \\
$=\{\mbox{Definition } \ref{Fder}\}$
\\
$F(V,\partial[\overline{V}] \{U.\{A.V\}_{k_{bs}}\}_{k_{bs}}) \sqcap $\\$F(V,\partial[\overline{V}] \{N_{B_{4}}^i.\{A.V\}_{k_{bs}}\}_{k_{bs}})$ \\
%%$=\{\mbox{Since } F=F_{MAX}^{EK}\}$\\
$=\{\mbox{Derivation in the definition } \ref{derivation}\}$
\\
$F(V,\{\{A.V\}_{k_{bs}}\}_{k_{bs}}) \sqcap F(V,\{N_{B_{4}}^i.\{A.V\}_{k_{bs}}\}_{k_{bs}})$ \\
$=\{\mbox{Since } F=F_{MAX}^{EK}\}$\\
$\{ A, B, S\}$(3.4)\\
\\
3- Compliance with the theorem \ref{PAT}:\\
\\
For any $U$, from (3.1) and (3.2) we have:\\
${\cal{W}}_{p,F}'(U,\{U.\{A.V\}_{k_{bs}}\}_{k_{bs}})=\{A, B, S\}\sqsupseteq \ulcorner U \urcorner \sqcap F(U,\partial  [\overline{U}]\{A.U.\{B.V\}_{k_{as}}\}_{k_{bs}} )=\ulcorner U \urcorner \cup \{A, B, S\}$ (3.5)\\
For any $V$, from (3.3) and (3.4) we have:\\
${\cal{W}}_{p,F}'(V,\{U.\{A.V\}_{k_{bs}}\}_{k_{bs}})=\{A, B, S\}\sqsupseteq \ulcorner V \urcorner \sqcap F(V,\partial  [\overline{V}]\{A.U.\{B.V\}_{k_{as}}\}_{k_{bs}} )=\ulcorner V \urcorner \cup \{A, B, S\}$ (3.6)\\
From (3.5)  and (3.6) we have: the messages exchanged in the session $S^{i}$  respect the theorem \ref{PAT} (III)

\section{Results and Interpretation}

The results of analysis of the amended version of the Woo-Lam protocol are summarized in Table \ref{WLAMGrowth}. From Table \ref{WLAMGrowth}, we conclude that this version fully respects the theorem \ref{PAT}. Hence, this protocol keeps its secrect inputs.

\begin{table}[h]
\caption{Compliance of the Woo-Lam protocol (amended version) with the Theorem \ref{PAT}}
\label{WLAMGrowth} 
   \centering
\scalebox{0.9}{%
\begin{tabular}{|c|c|c|c|c|c|}
  \hline
&  $\alpha$ &Role& $R^-$ & $r^+$ &  The.\ref{PAT}\\
  \hline
1&  $k_{ab}^i$ & $A$& $X$ & $\{B.k_{ab}^i\}_{k_{as}}$ & Ok\\
  \hline
2& $X$ &  $A$ & $X$ & $\{B.k_{ab}^i\}_{k_{as}}$& Ok\\
  \hline
 \hline
3&  $N_b^i$ & $B$& $A$ & $N_b^i$ & Ok\\
  \hline
4& $Y$ &  $B$ & $Y$ & $\{A.N_b^i.Y \}_{k_{bs}}$& Ok\\
  \hline
5& $N_b^i$ &  $B$ & $Y$ & $\{A.N_b^i.Y \}_{k_{bs}}$& Ok\\
  \hline
 \hline
6&  $U$ & $S$& $\{A.U.\{B.V\}_{k_{as}}\}_{k_{bs}}$ & $\{A.V\}_{k_{bs}}\}_{k_{bs}}$ & Ok\\
  \hline
7& $V$ &  $S$ & $\{A.U.\{B.V\}_{k_{as}}\}_{k_{bs}}$ & $\{U.\{A.V\}_{k_{bs}}\}_{k_{bs}}$&  Ok\\
 \hline
\end{tabular}
}
\\\hspace{\linewidth}

\end{table}

\section{Related Works}

Our witness-functions are comparable to the rank-functions of Steve Schneider~\cite{Schneider4}and the interpretation-functions of Houmani~\cite{Houmani1,Houmani3,Houmani8,Houmani5}. Unlike
the rank-functions, the witness-function are easy to build and easy to use. The rank-functions require CSP~\cite{Schneider96,SchneiderD04} and are difficult to search in a protocol~\cite{Shaikh1}. They could even not exist~\cite{Heather}.
Unlike the interpretation-functions, the witness-functions do not dictate that a message must be protected by the direct key. Any further protective key could define a witness-function. Our functions do not depend on variables thanks to their static bounds. That is a major fact. All that makes our witness-function more flexible and would allow us to prove correctness of a wider range of protocols.

\section{Conclusion and Future Work}

In this paper, we presented a new framework to analyze statically cryptographic protocols for secrecy using the witness-functions. We successfully tested them on an amended version of the Woo-Lam protocol. In a future work, we will test them on protocols with theories~\cite{cortier9900,cortier1971,cortier9901} and on compose protocols~\cite{cortier9903,cortier9902,cortier9904}. We believe that our witness-functions will help to treat these problems.

\bibliographystyle{unsrt}
\bibliography{Ma_these}

\begin{thebibliography}{10}

\bibitem{WitnessArt1}
Jaouhar Fattahi, Mohamed Mejri, and Hanane Houmani.
\newblock Secrecy by witness functions.
\newblock In {\em 5th Proceedings of the Formal Methods for Security Workshop
  co-located with the PetriNets-2014 Conference}, pages 34--52, 2014.

\bibitem{WitnessArt2}
Jaouhar Fattahi, Mohamed Mejri, and Hanane Houmani.
\newblock New functions for secrecy on real protocols.
\newblock In {\em Fourth International Conference on Computer Science,
  Engineering and Applications (ICCSEA 2014), {Chennai, India}}, pages
  229--250, 2014.

\bibitem{2014arXiv1408.2774F}
J.~{Fattahi}, M.~{Mejri}, and H.~{Houmani}.
\newblock {A Semi-Decidable Procedure for Secrecy in Cryptographic Protocols}.
\newblock {\em ArXiv e-prints}, August 2014.

\bibitem{WitnessArt3}
Jaouhar Fattahi, Mohamed Mejri, and Hanane Houmani.
\newblock Introduction to the witness-functions for secrecy in cryptographic
  protocols(inpress).
\newblock In {\em The 2014 International Conference on Networks and
  Information, {Nanjing, China}}, 2014.

\bibitem{Fatt1407:Relaxed}
Jaouhar Fattahi, Mohamed Mejri, and Hanane Houmani.
\newblock Relaxed {Conditions} for {Secrecy} in a {Role-Based} specification.
\newblock {\em International Journal of Information Security}, 1:33--36, July
  2014.

\bibitem{Debbabi11}
Mourad Debbabi, Y.~Legar{\'e}, and Mohamed Mejri.
\newblock An environment for the specification and analysis of cryptoprotocols.
\newblock In {\em ACSAC}, pages 321--332, 1998.

\bibitem{Debbabi22}
Mourad Debbabi, Mohamed Mejri, Nadia Tawbi, and I.~Yahmadi.
\newblock Formal automatic verification of authentication crytographic
  protocols.
\newblock In {\em ICFEM}, pages 50--59, 1997.

\bibitem{Debbabi33}
Mourad Debbabi, Mohamed Mejri, Nadia Tawbi, and I.~Yahmadi.
\newblock From protocol specifications to flaws and attack scenarios: An
  automatic and formal algorithm.
\newblock In {\em WETICE}, pages 256--262, 1997.

\bibitem{DolevY1}
Danny Dolev and Andrew Chi-Chih Yao.
\newblock On the security of public key protocols.
\newblock {\em IEEE Transactions on Information Theory}, 29(2):198--207, 1983.

\bibitem{Schneider4}
Steve Schneider.
\newblock Verifying authentication protocols in csp.
\newblock {\em IEEE Trans. Software Eng.}, 24(9):741--758, 1998.

\bibitem{Houmani1}
Hanane Houmani and Mohamed Mejri.
\newblock Practical and universal interpretation functions for secrecy.
\newblock In {\em SECRYPT}, pages 157--164, 2007.

\bibitem{Houmani3}
Hanane Houmani and Mohamed Mejri.
\newblock Ensuring the correctness of cryptographic protocols with respect to
  secrecy.
\newblock In {\em SECRYPT}, pages 184--189, 2008.

\bibitem{Houmani8}
Hanane Houmani and Mohamed Mejri.
\newblock Formal analysis of set and nsl protocols using the interpretation
  functions-based method.
\newblock {\em Journal Comp. Netw. and Communic.}, 2012, 2012.

\bibitem{Houmani5}
Hanane Houmani, Mohamed Mejri, and Hamido Fujita.
\newblock Secrecy of cryptographic protocols under equational theory.
\newblock {\em Knowl.-Based Syst.}, 22(3):160--173, 2009.

\bibitem{Schneider96}
Steve Schneider.
\newblock Security properties and csp.
\newblock In {\em IEEE Symposium on Security and Privacy}, pages 174--187,
  1996.

\bibitem{SchneiderD04}
Steve~A. Schneider and Rob Delicata.
\newblock Verifying security protocols: An application of csp.
\newblock In {\em 25 Years Communicating Sequential Processes}, pages 243--263,
  2004.

\bibitem{Shaikh1}
Siraj~A. Shaikh and Vicky~J. Bush.
\newblock Analysing the woo-lam protocol using csp and rank functions.
\newblock In {\em WOSIS}, pages 3--12, 2005.

\bibitem{Heather}
James Heather and Steve Schneider.
\newblock A decision procedure for the existence of a rank function.
\newblock {\em J. Comput. Secur.}, 13(2):317--344, March 2005.

\bibitem{cortier9900}
Hubert Comon-Lundh, V{\'e}ronique Cortier, and Eugen Zalinescu.
\newblock Deciding security properties for cryptographic protocols. application
  to key cycles.
\newblock {\em ACM Trans. Comput. Log.}, 11(2), 2010.

\bibitem{cortier1971}
V{\'e}ronique Cortier and St{\'e}phanie Delaune.
\newblock Decidability and combination results for two notions of knowledge in
  security protocols.
\newblock {\em J. Autom. Reasoning}, 48(4):441--487, 2012.

\bibitem{cortier9901}
V{\'e}ronique Cortier, Steve Kremer, and Bogdan Warinschi.
\newblock A survey of symbolic methods in computational analysis of
  cryptographic systems.
\newblock {\em J. Autom. Reasoning}, 46(3-4):225--259, 2011.

\bibitem{cortier9903}
Stefan Ciobaca and Veronique Cortier.
\newblock Protocol composition for arbitrary primitives.
\newblock {\em 2012 IEEE 25th Computer Security Foundations Symposium},
  0:322--336, 2010.

\bibitem{cortier9902}
V{\'e}ronique Cortier.
\newblock Secure composition of protocols.
\newblock In {\em TOSCA}, pages 29--32, 2011.

\bibitem{cortier9904}
V{\'e}ronique Cortier and St{\'e}phanie Delaune.
\newblock Safely composing security protocols.
\newblock {\em Formal Methods in System Design}, 34(1):1--36, 2009.

\end{thebibliography}


\begin{thebibliography}{1}

\bibitem{IEEEhowto:kopka}
H.~Kopka and P.~W. Daly, \emph{A Guide to \LaTeX}, 3rd~ed.\hskip 1em plus
0.5em minus 0.4em\relax Harlow, England: Addison-Wesley, 1999.

\end{thebibliography}

\begin{comment}

\end{comment}
%%\newpage
%%\Large{Appendix}
\normalsize
%%\include{WooLamfaille1}
%%\pagebreak
%%\include{WooLamfaille2}

\section*{Notice}
© 2014 IEEE. Personal use of this material is permitted. Permission from IEEE must be obtained for all other uses, in any current or future media, including reprinting/republishing this material for advertising or promotional purposes, creating new collective works, for resale or redistribution to servers or lists, or reuse of any copyrighted component of this work in other works.

%%\end{comment}
\end{document}